\documentclass[preprint]{aastex}%emulateapj}%

\usepackage{epsfig}
\usepackage{graphicx}

\shorttitle{Asymmetric Pupil Fourier Wavefront Sensor}
\shortauthors{Frantz Martinache}

\begin{document}

\title{The Asymmetric Pupil Fourier Wavefront Sensor}

\author{Frantz Martinache}
\affil{National Astronomical Observatory of Japan, Subaru Telescope,
  Hilo, HI 96720, USA}
\email{frantz@naoj.org}

\begin{abstract}
This paper introduces a novel wavefront sensing approach that
relies on the Fourier analysis of a single conventional direct
image. In the high Strehl ratio regime, the relation between the phase
measured in the Fourier plane and the wavefront errors in the pupil
can be linearized, as was shown in a previous work that introduced the
notion of generalized closure-phase, or kernel-phase. 
The technique, to be usable as presented requires two conditions to be
met: (1) the wavefront errors must be kept small (of the order of one
radian or less) and (2) the pupil must include some asymmetry, that
can be introduced with a mask, for the problem to become solvable.
Simulations show that this asymmetric pupil Fourier wavefront sensing
or APF-WFS technique can improve the Strehl ratio from 50 to over 90
\% in just a few iterations, with excellent photon noise sensitivity
properties, suggesting that on-sky close loop APF-WFS is possible with
an extreme adaptive optics system.
\end{abstract}


\section{Non-common path error and extreme AO}

Contrast limits for the direct imaging of extrasolar planets from
ground based adaptive optics (AO) observations are currently set by
the presence of static and slow-varying aberrations in the optical
path that leads to the science instrument \citep{2003EAS.....8..233M}.
Because some of these aberrations are not sensed by the wavefront
sensor, something called the non-common path error, they are
responsible for the presence of long lasting speckles in the image. 
Since extrasolar planets are faint unresolved sources, it is
impossible to discriminate them among these speckles in one single
frame. The family of differential imaging techniques is aimed at
calibrating out some of these static aberrations, in post processing
by using either sky rotation (angular differential imaging, or ADI),
polarization differential imaging (PDI), or wavelength dependence of
the speckles (spectral differential imaging or SDI).
Of these, ADI \citet{2006ApJ...641..556M} has been successful in most
notably producing the image of the planetary system around HR 8799
\citep{2008Sci...322.1348M}.

A new generation of high contrast instruments using an approach of
improved wavefront control called extreme adaptive optics (XAO), aims
at producing higher contrast raw images, by including additional high
density wavefront control devices, fed by advanced wavefront sensing
techniques to actively control the wavefront during the
observations. In all cases, the primary goal of the high order
deformable mirror is to calibrate the non-common path error between
the conventional AO system and the science camera, and then to try and
help creating a high contrast region in the image, by actively
modulating the speckles in the field, for instance using speckle
nulling \citep{2012arXiv1206.2996M}.

The control of the non common path error is an important element of
any XAO system, that requires the implementation of dedicated hardware
or special acquisition procedures: the Gemini Planet Imager
\citep{2010SPIE.7736E.179W} and the Project P1640
\citep{2012SPIE.8447E..6WZ} for instance, both include a dedicated
interferometric calibration unit, to characterize and compensate the
non-common path error.
Other options often involve some form of phase diversity, using
multiple acquisitions with the science camera while moving an internal
calibration source along the optical path with ESO's SPHERE
\citep{2010SPIE.7736E..13S} and JPL's PALM-3000
\citep{2010SPIE.7736E..58B} or even the science camera itself with
Subaru's SCExAO project \citep{2011SPIE.8151E..22M}.

This paper introduces an alternate approach to the calibration of the
non common path error in XAO systems, using a non-invasive
conventional hard-stop mask located in the pupil plane, and the direct
analysis of the Fourier properties of the resulting images acquired
with the focal plane camera.
The simulations presented in this work demonstrate remarkable
performance of the approach: combined with proper control of a
deformable mirror (DM) that is present in all XAO system, it can bring
the wavefront quality from a starting Strehl ratio of 50 \% up to over
95 \% in less than five iterations.
Because of the mask's very limited impact on the overall morphology of
the point spread function (PSF) and its high sensitivity, the
technique seems compatible with a general purpose imaging
(non-coronagraphic) instrument, and would enable continuous
tracking of the non-common path aberrations, to maintain very high
Strehl during observation.

\section{Wavefront Sensing in the Fourier-plane}
\label{s:fourier}

Even in near-perfect observing conditions, diffraction is a
significant hindrance to the interpretation of images at the highest
angular resolution. In this diffraction-limited regime, it is
advantageous to adopt an interferometric point of view of image
formation, and work not from images directly, but from their Fourier
Transform counterpart instead. 
Working on the Fourier Transform of images is often refered to as
working in the Fourier-plane, the spatial-frequency plane or the
uv-plane, the latter expression being widely used in sparse aperture
interferometry.
The Fourier Transform being a complex function, encodes information
both in terms of amplitude (or modulus) and phase: this paper focuses
solely on the phase.

To introduce the notion of kernel-phase, \citet{2010ApJ...724..464M}
proposes a simple but powerful description of how pupil phase errors
propagate into phase signal measured in the Fourier-plane, in terms of
linear algebra. This model, when it applies (when wavefront
errors are small i.e. $\varphi < 1$ radian for a conventional imaging
telescope), offers an interesting alternative to the traditional
object-image relation, written in terms of convolution. 
When this model applies, the (usually unknown) instrumental phase in
the pupil $\varphi$ can be related to the phases $\Phi$ measured in
the Fourier plane with a single, fully determined, linear operator
called the phase transfer matrix (noted $\mathbf{A}$).
The discrete model, presented in several references
\citep{2010ApJ...724..464M,   2011SPIE.8151E..33M,
  2012SPIE.8445E..04M} is briefly re-introduced here for convenience.

The aperture of any instrument, usually refered to as the pupil, can
be discretized into a finite collection of $n$ elementary
sub-apertures, using a regular grid (for instance square or hexagonal)
adapted to the actual pupil shape.
One of these elementary sub-apertures is taken as a zero-phase
reference: the pupil phase of a coherent point source is therefore
written as a $(n-1)$-component vector $\varphi$, measured relative to
this reference. 
Assuming that the image produced by the system is at least
Nyquist-sampled, in the Fourier-plane, one is able to sample up to $m$
phases, organized in a vector $\Phi$.

The complete model \citep{2010ApJ...724..464M}, also includes one
additional term $\Phi_{0}$, that encodes the true phase information
about the target of interest. The Fourier-phase $\Phi$ therefore
writes as:

\begin{equation} 
\Phi = \Phi_0 + \mathbf{A} \cdot \varphi.  
\label{eq:1}
\end{equation}

The motivation for this paper is to use this formalism not to learn
something about the target, which was shown to be recoverable using
kernel-phases, but to solve the wavefront sensing problem.
This paper will therefore from now on consider the target to be a
non-resolved source, so that $\Phi_0$ = 0. XAO systems typically
include a laser-fed single-mode fiber calibration source, for off-line
tests of wavefront control loops, that can be used in this purpose.

\begin{figure}
\plotone{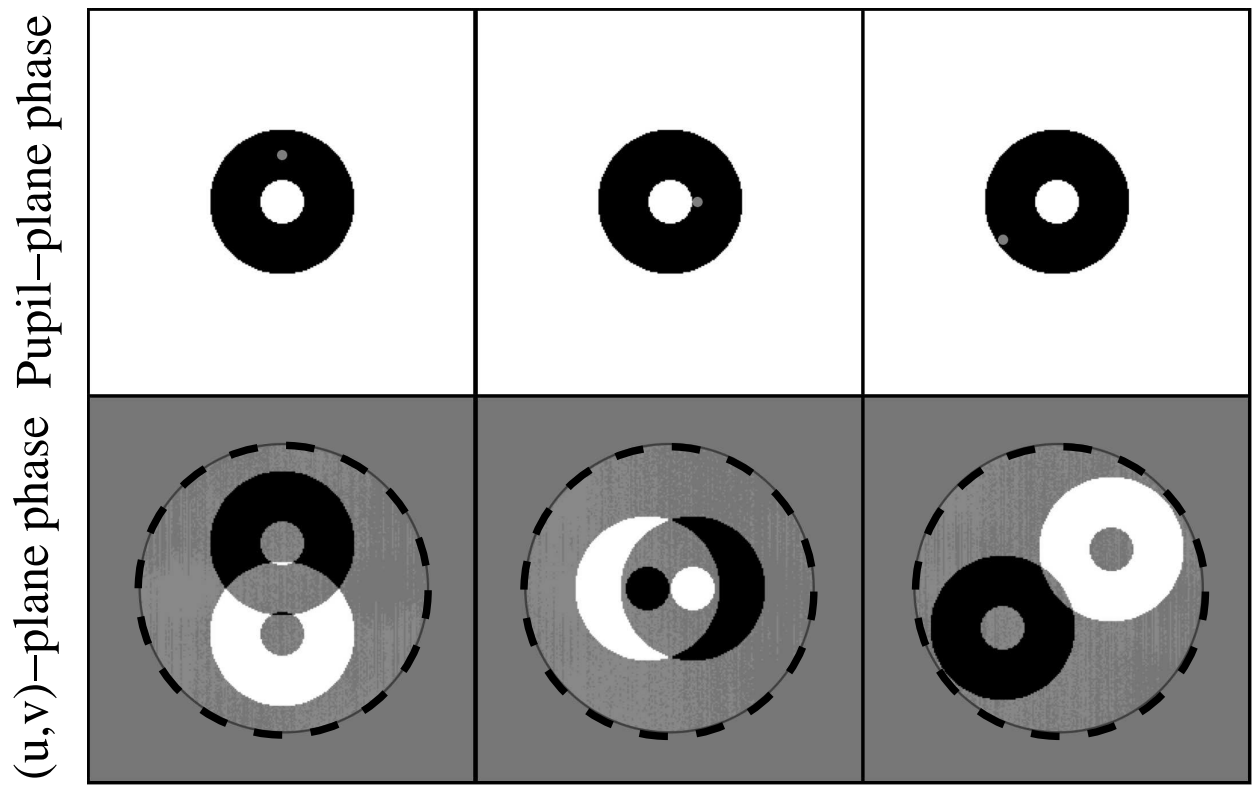}
\caption{
Illustration of the pupil phase propagation mechanism into
phase measured in the Fourier plane. The top row shows a pupil phase
``poke'' applied to three different sub-apertures of the full
two-dimensional pupil, in this case, a circular aperture with a 30 \%
central obscuration. The bottom row shows for each case the
corresponding distribution of phase that is observed in the Fourier
plane. The overlaid dashed-line circle in the bottom row marks the
cutoff spatial frequency of the transfer function.}
\label{f:poke}
\end{figure}

Wavefront sensing in the Fourier-plane is routinely achieved in long
baseline interferometry. However, to be able to handle the large
optical path differences (OPD) expected along a long baseline that can
be up to several hundred meters long in the optical, resulting in
phase shifts larger than $2\pi$, the strategy is to disperse the light
so as to increase the coherence length and solve the $2\pi$ wrapping
of the phase \citep{1921ApJ....53..249M, 1996ApOpt..35.3002K}.
This strategy, so far only used with a small number of apertures was
generalized by \citet{2004JOptA...6..216M}, for a all in one
Fizeau-type interferometer: an approach called ``dispersed-speckles''.
This approach appears as an relevant solution for the calibration of
OPD inside AO fed integral field spectrographs
\citep{2012arXiv1208.3190P}. In systems where the OPD is small 
($\sim$1 radian), dispersion isn't necessary, and the use of a single
wavelength may suffice in solving the problem.

This approach however has one major flaw: for a pupil like the one
shown in Fig. 1, it is insensitive to structures of the wavefront that
are even, that is for which $\varphi(-r) = \varphi(r)$. While an
analytical demonstration of this fundamental property is provided in
an appendix to the paper, one can build a more intuitive understanding
of wavefront sensing in the Fourier plane, by considering the
following scenarios, that illustrate how instrumental (pupil) phase
errors $\varphi$ propagate into phase measured in the Fourier plane:

\begin{itemize} 
\item If the phase is constant across the entire pupil, then none of
the baselines formed by any pair of elementary sub-apertures does
record a phase difference, and the phase in the Fourier plane is zero
everywhere: the non-common path error is perfectly calibrated.

\item If a phase offset $\delta_0$ is added to one single sub-element
of the aperture, something we will refer to a "poke", then each
baseline involving this sub-element records a phase difference, which
is exactly $\pm\delta_0$. Fig. 1 presents several such scenarios.

\item If the pupil-plane phase error $\varphi$ is completely random,
each of the samples in the Fourier plane is then the average of
$\mathbf{R}$ phase differences on the pupil, where $\mathbf{R}$ is a
vector summarizing the redundancy of the baselines identified in the
system.
\end{itemize} 

To see that this approach to wavefront sensing would be insensitive to
an even aberration, the reader can consider either of the scenarios
shown in Fig. \ref{f:poke}, and observe that the phase pattern
recorded in the uv-plane for either of these ``pokes'' is perfectly
odd: $\Phi(-r) = - \Phi(r)$, a property expected to the fact that it
results from the Fourier Transform of a real image.

\begin{figure} 
\plotone{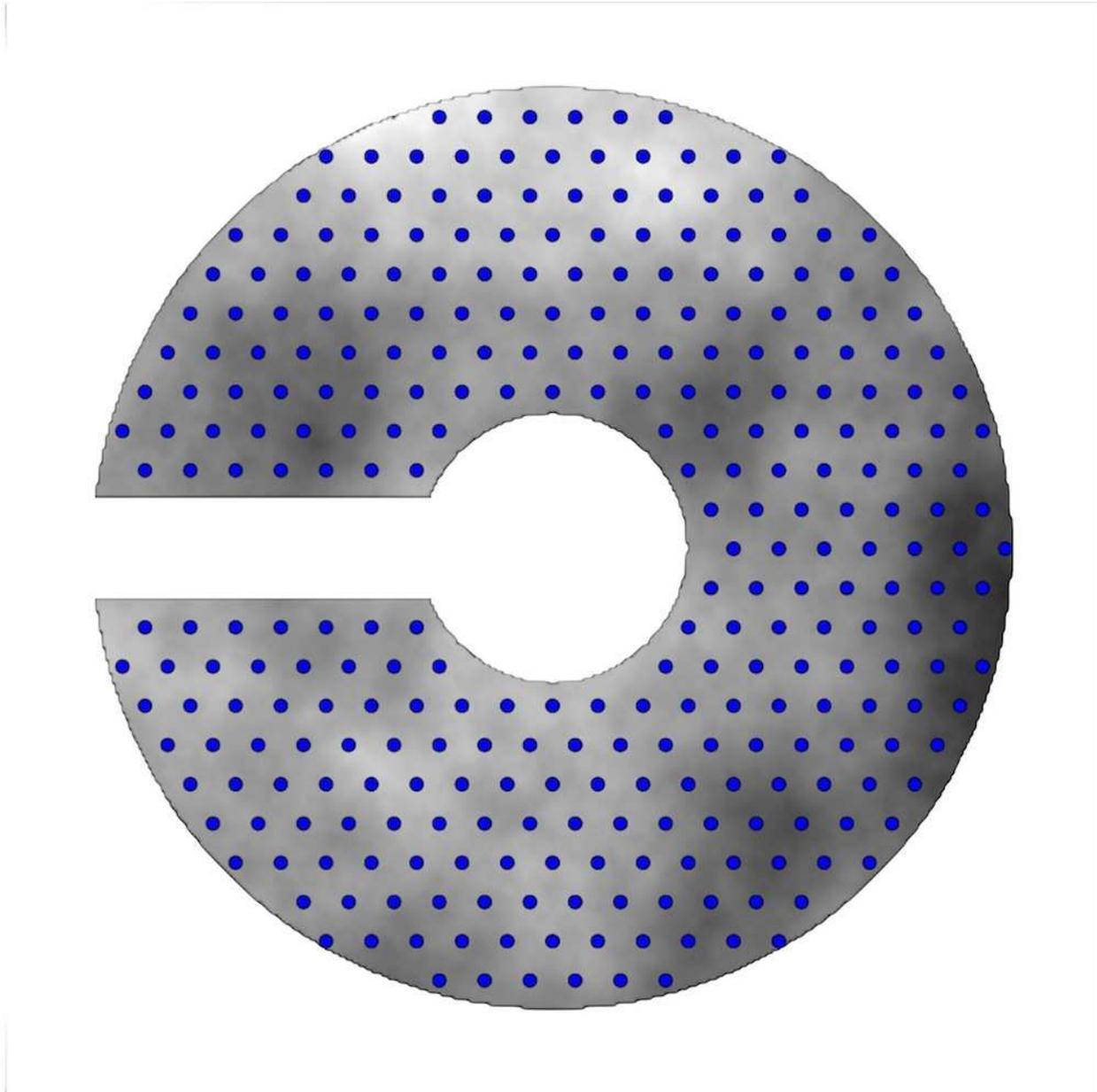}
\caption{ 
Discrete hexagonal model used to construct the phase transfer matrix
$\mathbf{A}$. Each of the thick dots overlaying the "true" image of
the input pupil phase inside the asymmetric pupil, marks the location
of a discrete pupil sample. This model contains 309 distinct pupil
sample points.
}
\label{f:asym}
\end{figure}

If the same amplitude ``poke'' $\delta_0$ is applied to the
diametrically opposed sub-aperture in the same direction (therefore
creating an elementary even aberration), the two resulting phase
patterns in the Fourier-plane will simply add to zero, and give the
illusion that that the system is perfectly in phase, no matter what
the amplitude of the ``poke''.
One equivalent way to say it is that such a Fourier-based wavefront
sensor is only sensitive to the odd component of the input wavefront.

One simple way to circumvent this severe limitation is to introduce some
asymmetry in the system, and the simplest thing to alter is the
pupil. High contrast instruments are usually equiped with a multi-slot
pupil wheel (ie. for coronagraphic Lyot-stops), and for this work, we
propose to modify the pupil used in Fig. \ref{f:poke} by adding one
single fairly thick spider arm that connects the outer edge of the
pupil to the edge of the central obstruction. The actual pupil used
for the results presented in this work is shown in Fig. \ref{f:asym}.

\section{Asymmetric Pupil Fourier Wavefront Sensor} 
\label{s:principle}

We need to go back to the instrumental phase propagation model given
summarized by eq. \ref{eq:1}, to see how the asymmetry introduced in
the pupil affects the properties of the phase transfer matrix
$\mathbf{A}$. 
To build the actual matrix, the pupil is modeled into a discrete grid
of sub-apertures that follow a regular hexagonal grid. Other options
(eg. square grid) remain possible, and have been used so far for
kernel-phase analysis \citep{2010ApJ...724..464M,2012pope}. 
The density of the grid should in practice be matched to the density
of actuators the deformable mirror that controls the incoming
wavefront. The image is expected to be at least Nyquist-sampled.

Without the missing sub-apertures along the thick arm, the model would
contain $n_a = 330$ pupil samples, propagating into $n_{uv} = 708$
distinct samples in the Fourier-plane. 
Just like for the kernel-phase, the proposed analysis relies on the
spectral properties of the phase transfer matrix $\mathbf{A}$. The
singular value decomposition (SVD) of the $708 \times 330$ resulting
phase transfer matrix $\mathbf{A}$, given by:

\begin{equation}
\mathbf{A} = \mathbf{U} \cdot \mathbf{\Sigma} \cdot \mathbf{V^T},
\label{eq:svd}
\end{equation}

\noindent
where $\mathbf{U}$ and $\mathbf{V^T}$ are unitary matrices
and $\mathbf{\Sigma}$ is a diagonal matrix containing the singular
values of $\mathbf{A}$, reveals that for this symmetrical geometry,
165 (that is exactly $n_a/2$) of the singular values of $\mathbf{A}$
are non-zero. Using this knowledge, it therefore appears possible to
build a pseudo-inverse for $\mathbf{A}$ that will enable the recovery
of one half of the wavefront phase in the input pupil. This
observation made after examination of the singular values of the phase
transfer matrix, echoes the conclusion made in
Sec. \ref{s:fourier}, that a Fourier based WFS can only sense the
odd component of the input wavefront, in terms of linear algebra. 

With the asymmetric pupil presented in Fig. \ref{f:asym}, things
apparently change very little: the number of apertures in the pupil
may drop to 309 (a loss of 21 apertures), they still project onto the
same space of 708 samples in the Fourier-plane, because of the
redundance.
However, a major change occurs in terms of spectral properties of this
new phase transfer matrix. The SVD of the now $708 \times 308$ linear
operator reveals that all the required 308 singular values are now
non-zero. A pseudo-inverse for $\mathbf{A}$ for such a configuration
therefore enables full recovery of both odd and even parts of the
wavefront.

In practice, the approach is used as follows: assuming that the
residual aberrations on the wavefront are of the order of one radian
or less (RMS), one asymmetric mask such as the one shown in
Fig. \ref{f:asym} is introduced in the pupil of the instrument,
before the final focal plane, while shining a calibration source
(often a laser-fed single mode fiber). With nothing else being
actuated, one (or more) image is acquired with the science detector,
which will be assumed to be Nyquist sampled.

The data reduction procedure is very similar to what has been
described for non-redundant masking interferometry data
\citep{2008ApJ...679..762K} and kernel-phase data analysis
\citep{2010ApJ...724..464M}. 
Each image is simply recentered, and and multiplied by a window
function to limit sensitivity to readout noise and filter out the
spatial frequency content of the image that is beyond the control
region of the wavefront control device.
A Fourier transform for each image is then calculated, and the phase
of this Fourier transform is sampled according to the discrete model
described in Section \ref{s:fourier}.
The resulting vector $\Phi$ is then simply projected back onto pupil
phase $\varphi$ inside the region unocculted by the mask, using the
pseudo inverse of $\mathbf{A}$.

\section{APF-WFS performance}

\subsection{SVD modes for the APF-WFS}

The wavefront reconstruction by the APF-WFS relies upon the
determination of singular value modes of the phase transfer matrix 
that in turn solely depend on the geometry of the pupil sampling. The
modes, which form an orthonormal basis for the wavefront, are also
sorted in order of decreasing level of significance, proportional to
their associated singular value.

\begin{figure}
\plotone{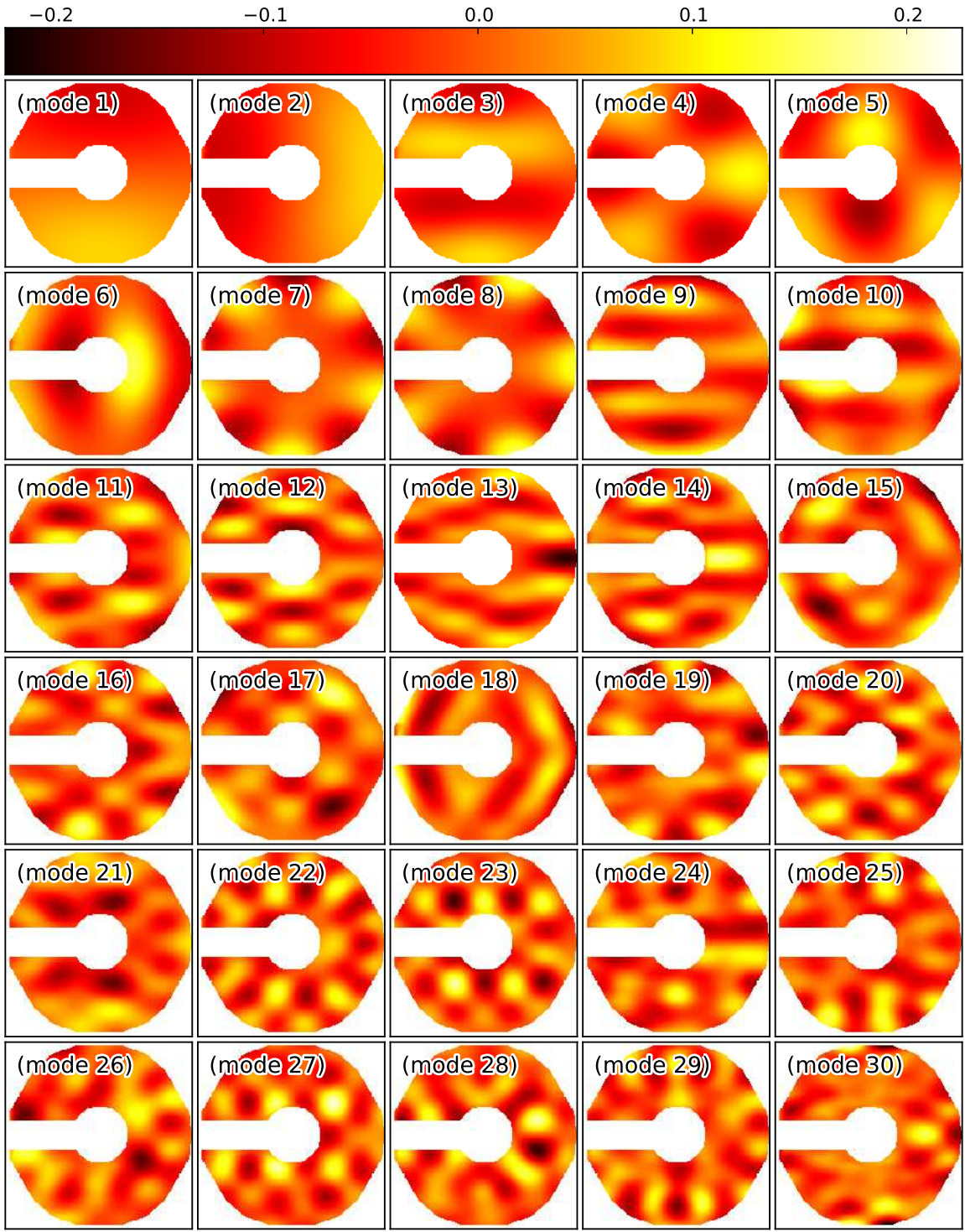}
\caption{
Representation of the first 30 SVD modes for the asymmetric pupil
geometry shown in Fig. \ref{f:asym}, sorted in order of decreasing
significance (proportional to their associated singular value). All
modes are represented using a common color scale, that spans the
$\pm$0.2 radian range.
}
\label{f:modes}
\end{figure}

Fig. \ref{f:modes} shows the first 30 modes (out of the total 309) the
SVD of the phase transfer matrix produces. One will observe that
although they do not closely match a familiar pattern of modes like
the Zernike polynomials or the series of sines and cosines, the lower
spatial frequencies (slope and curvature) are naturally reconstructed
first, and higher spatial frequency content appears as the order of
the modes increases, guided by the hexagonal geometry chosen to model
the pupil.

\subsection{Optimization of the pseudo inverse for wavefront
  reconstruction}

\begin{figure}
\plotone{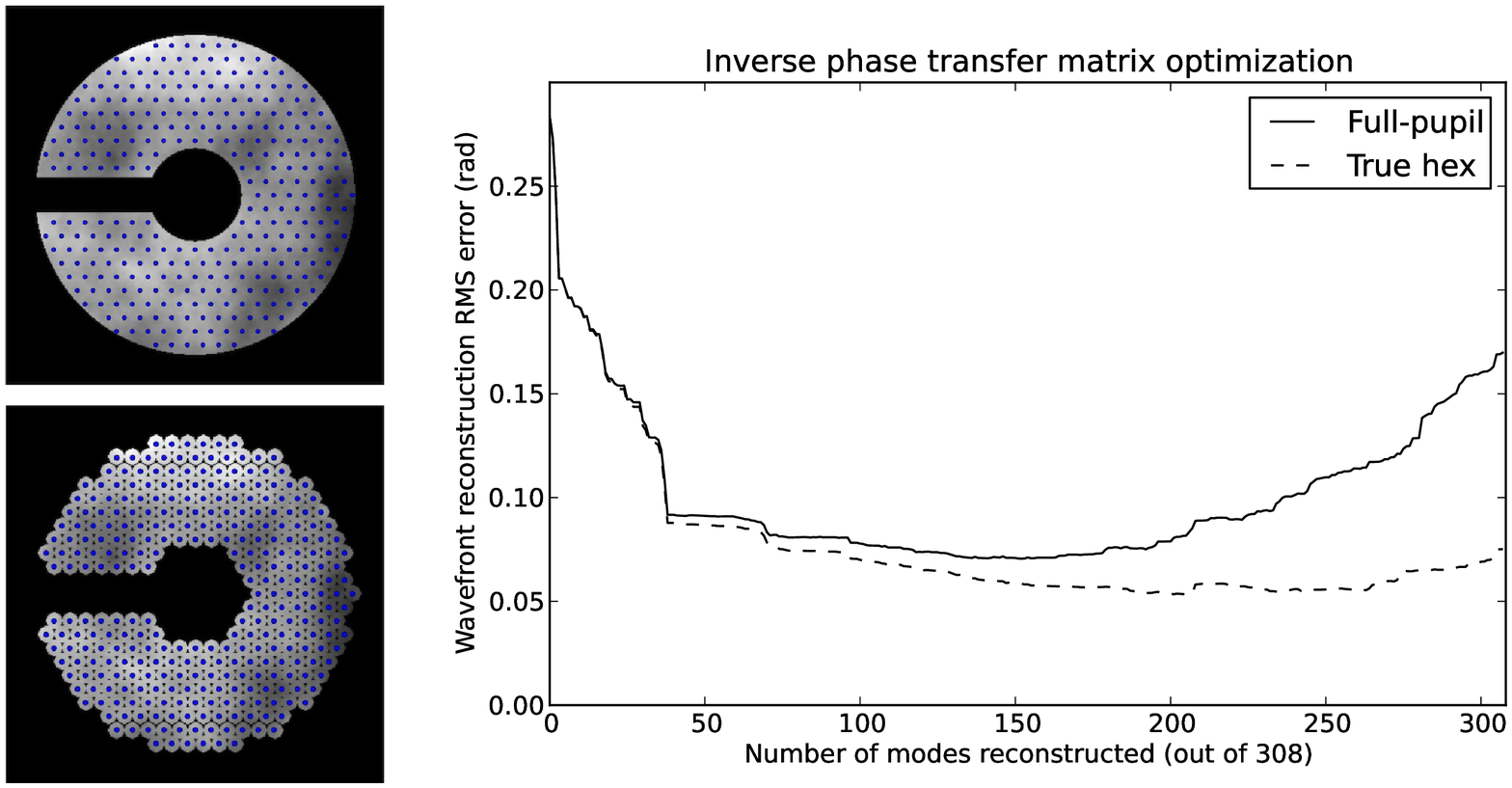}
\caption{
Curve of the APF-WFS wavefront reconstruction error
(RMS) as a function of the number of modes included in the
reconstruction. Two curves are shown to highlight the impact of the
discrete representation of the wavefront. The solid line curve, shows
the reconstruction RMS for the circular pupil shown in
Fig. \ref{f:asym}, and reproduced in the top left panel. 
The figure shows that including the first 40 modes exhibits the most
important increase in the wavefront reconstruction fidelity, which
then slowly increases until reaching an optimum around 150
modes. Including more modes however increases the reconstruction
error. The dashed line curve shows how the reconstruction RMS evolves
for the labeled "true hex" pupil (bottom left panel) that more closely
matches the discrete sampling used to model the optical system.
}
\label{f:optim}
\end{figure}

While fairly dense, the 309-element discrete representation of the
pupil used to establish the linear relation for this APF-WFS remains
an approximate representation of the continuous circular aperture that
is the real pupil. While not significant for low-order modes of the
wavefront, the impact of this discrete sampling is expected to
increase for the high order modes. This, combined with the fact that
the same high order modes, are associated to small singular values
(for this geometry, the singular values span over three orders of
magnitude from $\sim$0.1 for the lowest to $\sim$150 for the highest),
suggests that the wavefront reconstruction fidelity by APF-WFS will
depend on the total fraction of modes one tries to reconstruct.
The result of a series of simulations, presented here, involving the
reconstruction of an identical input wavefront using different
reconstructions illustrates this property.

The input wavefront is a Kolmogorov phase screen, scaled down to a
Strehl ratio $\sim$ 85 \%. One asymmetric pupil mask, such as the one
shown in Fig. \ref{f:asym} is used to simulate the acquisition of one
single, photon noise free PSF, processed following the procedure
highlighted in Sec. \ref{s:principle}.
We look at the result of a reconstruction of the input wavefront,
based on the direct inversion of eq. \ref{eq:1}:

\begin{equation}
  \varphi = \mathbf{A_k^+} \cdot \Phi,
\label{eq:inv}
\end{equation}

\noindent
where $\mathbf{A_k^+}$ is one possible pseudo-inverse of the phase
transfer matrix $\mathbf{A}$, calculated only after keeping the first
$k$ singular values contained in $\mathbf{\Sigma}$:

\begin{equation}
  \mathbf{A_k^+} = \mathbf{V} \cdot \mathbf{\Sigma_k^+} \cdot
  \mathbf{U^T}.
\end{equation}

Each of these pseudo inverses then acts upon the same sampled Fourier
phase vector $\Phi$ to produce an estimate of the
wavefront. Fig. \ref{f:optim} plots the reconstruction error RMS as a
function of the number of modes used for the pseudo inverse of the
phase transfer matrix, using anything between one and 309 modes (the
maximum for this pupil model).
The solid-line curve shows the evolution of the reconstruction error for
the a simulated circular aperture like shown in Fig. \ref{f:asym}. 
One will observe that the error exhibits its most rapid improvement
when the number of modes included in the reconstruction increases from
1 to about 40. This is somewhat expected, given that a Kolmogorov
wavefront exhibits more power in the low-spatial frequencies.
The reconstruction keeps slowly improving and reaches an optimum when
150 modes are used. Beyond that, the wavefront reconstruction
residuals eventually increase, and the performance of the
reconstruction degrades. 
With noise-free data, even the low-significance singular values
corresponding to high spatial frequency modes should be well
reconstructed, so this reduction of the wavefront reconstruction
fidelity requires another explanation: the unperfect match between the
hexagonal grid sampling and the actual pupil geometry, to explain this
behavior.
To confirm  this hypothesis, Fig. \ref{f:optim} also shows the
residuals obtained under identical conditions of wavefront and signal
to noise, with a pupil mask, labeled ``true hex'', that more closely
matches the sampling used in the linear model: while the inclusion of
the first 40 modes shows the same increase as with the circular pupil,
the reconstruction fidelity keeps improving as the number of included
modes increases. 
While an optimum is nevertheless observed (around 200 modes), the
inclusion of all 309 available modes in the reconstruction does not
significantly degrade the performance. The RMS can in fact be brought
abitrarily close to zero by reducing the size of the sub-apertures of
the ``true hex'' model or equivalently, by filtering out more
aggressively the spatial frequency content beyond the sampling.

While a pupil mask with this ``true hex'' geometry can easily be
manufactured, one may just as easily cknowledge this characteristic as
a feature of the APF-WFS, and deliberately choose to adjust the total
number of modes used in the reconstruction to the optimum found for a
simulation in the noise-free case.  The close loop results shown in
the next section will therefore be limited to the reconstruction of
the first 150 modes.

\subsection{Steady state close-loop simulation}

\begin{figure*}
\plotone{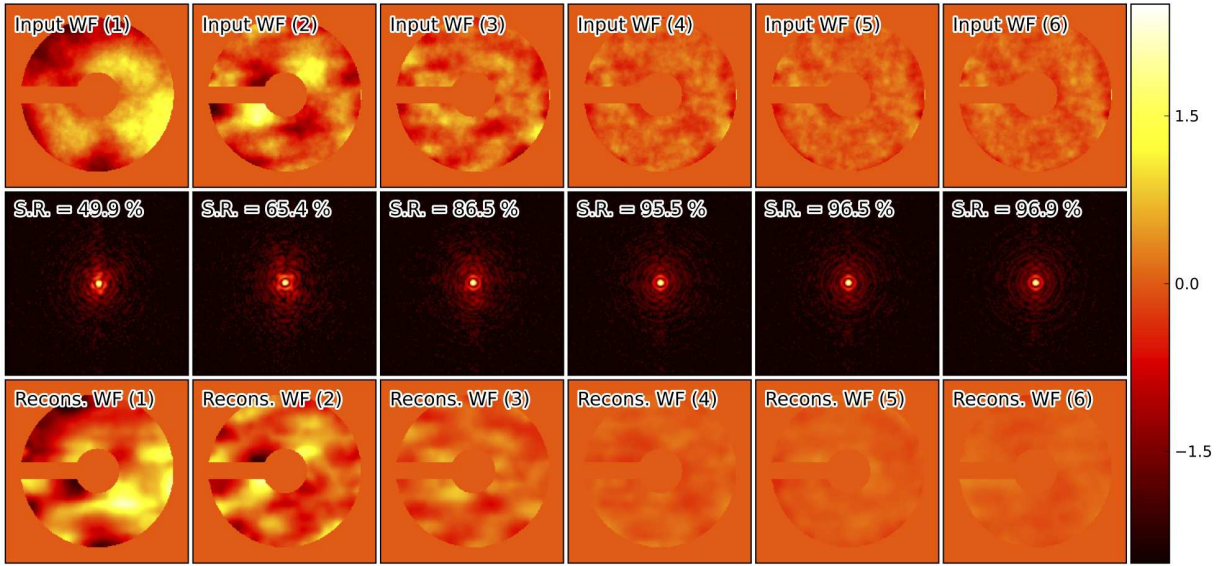}
\caption{
Sequence of wavefronts and resulting PSFs in a close-loop simulation
of the APF-WFS, for 150 modes. The starting point of the simulation is
a static Kolmogorov-type wavefront with a RMS 0.8 radians, resulting
in a 50 \% Strehl PSF. From left to right, the figure shows the impact
of successive iterations of the APF-WFS. The top row shows the
simulated wavefront used to produce the (photon-noise limited) PSFs
shown in the middle row. The bottom row shows the wavefront
reconstructed from the analysis of the PSFs by the APF-WFS. All the
wavefronts are represented using a common color scale, that spans the
$\pm$2 radian range.
}
\label{f:cloop}
\end{figure*}

One anticipated application of this technique is the characterization
of the small non-common path error between the fast wavefront sensor
arm and the science camera in an XAO system. This simulation
illustrates this use case of the technique. After the initial close
loop on the high order wavefront sensor, the Strehl ratio on the
science focal image is of the order of 50 \%, that corresponds to a
wavefront RMS error of 0.8 radians.

For each iteration, a photon noise limited ($10^5$ photons for this
series of images) single PSF is acquired, and the wavefront
reconstructed by the APF-WFS is then fed back to the upstream high
order wavefront sensor to offset its ``flat'' shape.

The simulation does not include any modeling of the deformable mirror,
and simply assumes that the wavefront determined by the APF-WFS can be
perfectly accounted for by the wavefront control system.
Fig. \ref{f:cloop} shows the result of five consecutive iterations of
one such close-loop simulation, reconstructing the first 150 SVD modes
of the wavefront each time. 
Three iterations are sufficient to bring the original 50 \% Strehl
ratio PSF to a steady Strehl level better than 95 \%. Given that the
linear model holds as long as $\varphi < 1$ radian, the method will
converge with an initial Strehl as low as $\sim 35$ \%.

\subsection{Sensitivity}

In his overview of the limits of adaptive optics,
\citet{2005ApJ...629..592G} defines a parameter noted $\beta_p$ that
characterizes the sensitivity to photon noise of a wavefront sensor.
The RMS reconstruction error by a wavefront sensor $\sigma_m$ for a
single mode due to photon noise is equal to:

\begin{equation}
\sigma_m = \beta_p \sqrt{\frac{1}{n_{ph}}},
\label{eq:betap}
\end{equation}

\noindent
where $n_{ph}$ represents the total number of photons used by the
sensor to reconstruct this mode. According to
\citet{2005ApJ...629..592G}, the ideal wavefront sensor exhibits a
photon noise sensitivity criterion $\beta_p = 1$, for each mode
sensed.
For reference, one of the best known and widely used wavefront sensing
techniques: the Shack-Hartmann, exhibits a very non-uniform
sensitivity over the modes it covers, with a peak sensitivity location
that depends on the number of elements in the Hartmann screen. It is
especially poorly sensitive ($\beta_p >> 1$) for lower modes.

\begin{figure}
\plotone{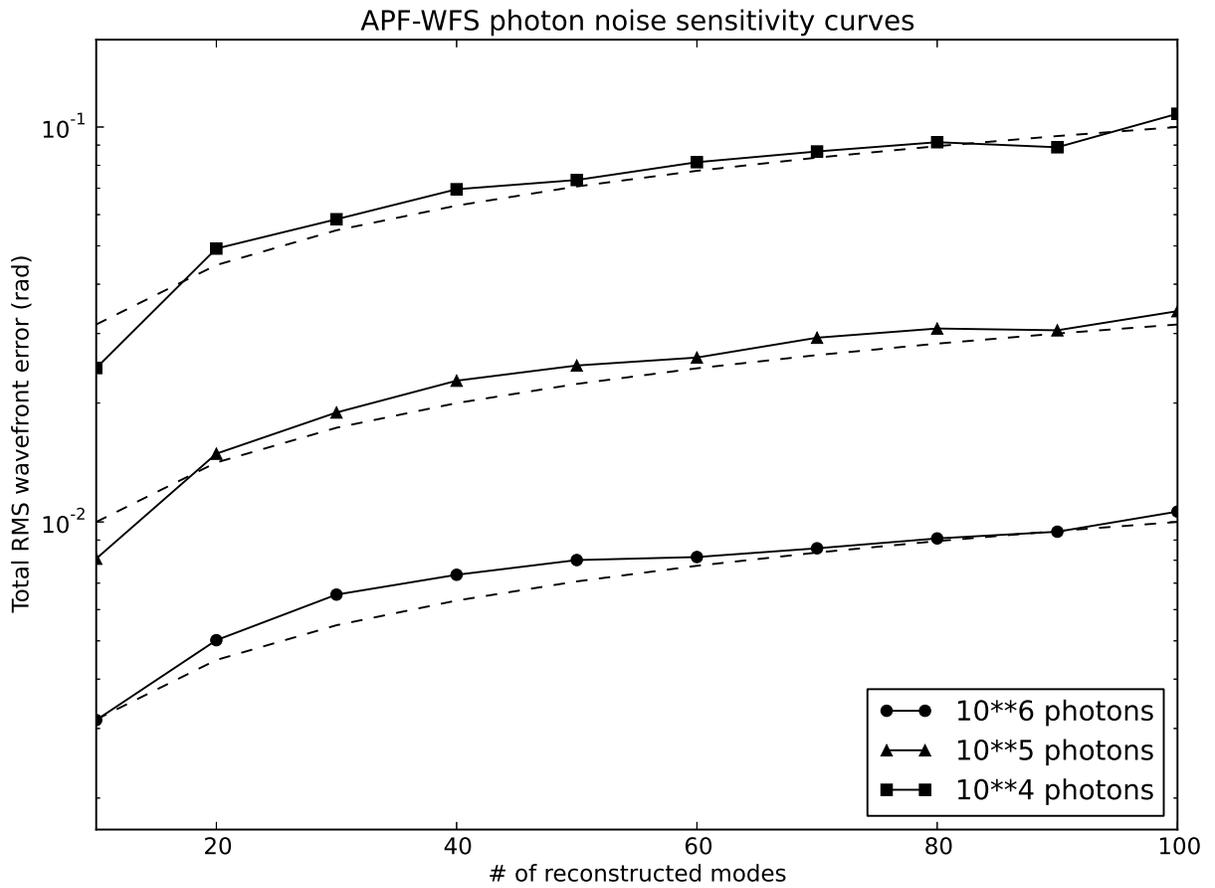}
\caption{
Sensitivity study of the APF-WFS. For three levels of total
illumination (10$^4$, 10$^5$ and 10$^6$ photons), the figure shows the
evolution of the total RMS wavefront reconstruction error (in radians)
due to photon noise.
}
\label{f:sensy}
\end{figure}

To illustrate how close to the ideal wavefront sensor the APF-WFS
approach is, Fig. \ref{f:sensy} plots three curves that show the
evolution of the RMS wavefront reconstruction error for different
level of illumination (10$^4$, 10$^5$ and 10$^6$ photons), for a
random wavefront with a Kolmogorov-type structure and a Strehl ratio
of 85 \%. The three curves, in this linear-log plot, equidistant in
the vertical direction, show that the overall performance 
follows the expected $1/\sqrt{n_{ph}}$ trend
(cf. eq. \ref{eq:betap}).

For each of these series of simulations, Fig. \ref{f:sensy} also
plots a dashed line, that represents the behavior of the ideal
wavefront sensor described by \citet{2005ApJ...629..592G}, with
constant and optimal sensitivity $\beta_p=1$ for all modes:

\begin{equation}
\sigma = \sqrt{\frac{n_m}{n_{ph}}},
%\label{eq:betap}
\end{equation}

\noindent
where $n_m$ and $\sigma$ now respectively represent the total number
of modes included in the reconstruction, and the total RMS wavefront
error in radians.
The close match between the three solid lines and the dashed lines
over the wide range of number of modes covered (from 10 to 100) is
remarkable and demonstrates that the APF-WFS, in the high-Strehl
regime where it is expected to operate, exhibits almost optimal
sensitivity, making it a very attractive option, in comparison to more
classical wavefront sensors.

\section{Conclusion}

This presentation of the asymmetric pupil Fourier wavefront sensor
completes the theoretical study of the linear model first used to
introduce the notion of kernel-phase. The linear model, although only
valid in the high Strehl regime, is very powerful. The eigen vectors
corresponding to zero singular values lead to kernel-phases, that in
turn encode information about the target of interest, and the eigen
vectors corresponding to non-zero singular values provide information
about the wavefront.

To recover the wavefront, some asymmetry needs to be introduced in the
pupil, and one (or more) mask such as the one used for this work is a
simple addition with a minimal impact, that most XAO instruments in
the making should be able to afford and accomodate. Using this mask
alone and a calibration source indeed enables the characterization of
the wavefront from the final science detector in an accurate and very
efficient manner. The region of the pupil hidden by the asymmetric
mask is obviously not accounted for, and for a complete
characterization of the wavefront over the entire pupil, serial
introduction of a pair of masks with asymmetric arms at different
azimuths is required.

Some improvements to this approach can already be anticipated: the
discrete grid model may be refined by simply (1) increasing its
density, and (2) including coefficients to take into account the fact
that some sample points are on the edge of the pupil. 
The latter point may become an essential addition if one wants to make
the technique compatible with apodized pupils, which are required for
high contrast coronagraphy.

Given its appealing photon noise sensitivity properties, and the
relatively weak impact on the overall PSF morphology, a close loop
on-sky operation appears as a viable possibility, for
non-coronagraphic observations. In rich fields, well separated PSFs
could simultaneously be used and the corresponding estimates of the
wavefront averaged to increase the signal to noise ratio.
The technique as it is, focuses on the phase: wavefront amplitude
errors which would require a dedicated simultaneous image of the
pupil, are going to be interpreted as phase aberrations.
If such amplitude errors are significant, or if the field of the
instrument extends beyond the anisoplanatic patch, the technique,
looking at multiple sources may however be able to do a more complete
tomographic reconstruction of the wavefront in a high Strehl MCAO
system, and identify the origin of amplitude aberrations.

All simulations done so far were monochromatic, which is acceptable
given that the primary application is the characterization of
non-common path error in an XAO system using an internal calibration
source.
But an on-sky close-loop system will require operation in a broader
band, and future work should explore the impact of broad band on
general performance of the wavefront sensor. While one expects it to
degrade somewhat, positive experience with kernel-phase analysis of
data acquired in broad band filters (up to 40 \%) demonstrates that
the linear model still holds, suggesting that the proposed approach
should remain valid.

The technique is currently being implemented on the SCExAO system,
with two masks with the asymmetric arm at different position angles,
so as to be able to recover the wavefront over the entire pupil. 
A more detailed description of this implementation along with an
experimental characterization of the sensor's performance will be the
object of a future publication.

\acknowledgements{The author thanks Olivier Guyon for the useful
  discussions of the ideas presented in this work.}

\appendix{}
\label{s:appen}

This section demonstrates how in the absence of pupil asymmetry, even
modes (for which $\varphi(-r) = \varphi(r)$) cannot be sensed in the
Fourier plane. Classical diffraction theory tells us that the instant
PSF is the square modulus of the Fourier Transform of the complex
amplitude in the pupil:

\begin{equation}
I = || \mathcal{F}(Ae^{i\varphi}) ||^2,
\end{equation}

\noindent
where $A$ and $\varphi$ respectively stand for the amplitude and phase
of the wavefront in the pupil, and $\mathcal{F}$ symbolizes the
Fourier Transform. The so-called Fourier- or uv-plane, as it is used
in the paper, is the Fourier Transform of the image, and equivalent
to the autocorrelation of the complex amplitude in the pupil plane, by
virtue of the convolution theorem.

\begin{equation}
  \mathcal{A} = Ae^{i\varphi} \otimes Ae^{i\varphi}.
  \label{eq:ac}
\end{equation}

In general, Kernel-phase and the wavefront sensing idea presented in
this paper rely on the assumption that wavefront errors are small, so
that the expression for the complex amplitude can be linearized:
$Ae^{i\varphi} \approx A (1 + i \varphi)$.
To simplify the notations, the functions describing the amplitude and
phase in the pupil will be written as functions of a single variable
$x$. Using the linearization, and eliminating second order terms, the
convolution product of eq. \ref{eq:ac} can be explicided as follows:

\begin{eqnarray}
\mathcal{A}(u) &\approx&
\int_{-\infty}^{+\infty}A(x)A(x+u)(1+i\varphi(x)+i\varphi(x+u))\, dx
\\
&=& 
\int_{-\infty}^{+\infty}A(x)A(x+u) \, dx + 
i \int_{-\infty}^{+\infty}A(x)A(x+u)\varphi(x+u)\, dx + 
i \int_{-\infty}^{+\infty}A(x)A(x+u)\varphi(x)\, dx.
\label{eq:expli}
\end{eqnarray}

The first (real) term of eq. \ref{eq:expli} only depends on the
function describing the pupil amplitude $A(x)$, and not on the
wavefront error $\varphi(x)$. 
We will therefore from now on only consider the imaginary part of this
autocorrelation. To further simplify the equations, we introduce one
new function $B(x) = A(x)\varphi(x)$. The imaginary part of this
autocorrelation can be written as:

\begin{equation}
\Im(\mathcal{A}(u)) = \int_{-\infty}^{+\infty} A(x) B(x+u) \, dx + 
\int_{-\infty}^{+\infty} A(x+u) B(x) \, dx.
\label{eq:integ1}
\end{equation}

A variable change $x' = x + u$ for the second integral term of
eq. \ref{eq:integ1} allows to rewrite the imaginary part of the
autocorrelation as:

\begin{equation}
\Im(\mathcal{A}(u)) = \int_{-\infty}^{+\infty} A(x) B(x+u) \, dx + 
\int_{-\infty}^{+\infty} A(x) B(x-u) \, dx.
\label{eq:integ2}
\end{equation}

The infinte integration domain can be split into two semi-infinite
domains:

\begin{equation}
\Im(\mathcal{A}(u)) = 
\int_{-\infty}^{0} A(x) B(x+u) \, dx + 
\int_{0}^{+\infty} A(x) B(x+u) \, dx + 
\int_{-\infty}^{0} A(x) B(x-u) \, dx + 
\int_{0}^{+\infty} A(x) B(x-u) \, dx,
\label{eq:integ2}
\end{equation}

\noindent
and one additional variable change $x' = -x$ is used for the integral
terms over the negative domains. Eq. \ref{eq:integ2} can be rewritten
as:

\begin{equation}
\Im(\mathcal{A}(u)) = -
\int_{0}^{+\infty} A(-x) B(-x+u) \, dx + 
\int_{0}^{+\infty} A(x)  B(x+u)  \, dx - 
\int_{0}^{+\infty} A(-x) B(-x-u) \, dx + 
\int_{0}^{+\infty} A(x)  B(x-u) \, dx.
\label{eq:integ3}
\end{equation}

The pupil is assumed to be symmetric, so that the amplitude function
verifies $A(-x) = A(x)$. Similarly, the even component of the
pupil phase also verifies $\varphi(-x) = \varphi(x)$. For
eq. \ref{eq:integ3}, this translates into $B(-x) = B(x)$.
For an even mode, the imaginary part of the autocorrelation becomes:

\begin{equation}
\Im(\mathcal{A}(u)) = -
\int_{0}^{+\infty} A(x)  B(x-u) \, dx + 
\int_{0}^{+\infty} A(x)  B(x+u)  \, dx - 
\int_{0}^{+\infty} A(x)  B(x+u) \, dx + 
\int_{0}^{+\infty} A(x)  B(x-u) \, dx.
\label{eq:even}
\end{equation}

The reader will observe that all terms in eq. \ref{eq:even} cancel
out, which demonstrates that even modes of the pupil wavefront cannot
be sensed in the Fourier plane. By contrast, the antisymmetric
properties of an odd mode translates into $B(-x) = - B(x)$. 
Two of the four terms cancel out, leading to the simplified expression
for the imaginary part of the autocorrelation:

\begin{equation}
\Im(\mathcal{A}(u)) = 2 \int_{0}^{+\infty} A(x) B(x+u) \, dx.
\label{eq:odd}
\end{equation}

%\bibliographystyle{../../../aa}
%\bibliography{../../../ms}

\end{document}